\documentclass{article}

\usepackage{arxiv}

\usepackage[utf8]{inputenc} 
\usepackage[T1]{fontenc}    
\usepackage{url}            
\usepackage{booktabs}       
\usepackage{amsfonts}       
\usepackage{nicefrac}       
\usepackage{microtype}      
\usepackage{lipsum}		
\usepackage{amsmath}
\usepackage{graphicx}
\usepackage{natbib}
\usepackage{doi}

\graphicspath{{Images/}}

\title{A Bayesian hierarchical model for improving exercise rehabilitation in mechanically ventilated ICU patients}

\author{{\hspace{1mm}Luke Hardcastle}\thanks{Corresponding author} \\
	Department of Statistical Science\\
	University College London\\
	London, UK \\
	\texttt{luke.hardcastle.20@ucl.ac.uk} \\
	\And
	{\hspace{1mm}Samuel Livingstone} \\
	Department of Statistical Science\\
	University College London\\
	London, UK \\
	\And
	{\hspace{1mm}Claire Black} \\
	Department of Statistical Science\\
	University College London\\
	London, UK \\
	\And
	{\hspace{1mm}Federico Ricciardi} \\
	Department of Statistical Science\\
	University College London\\
	London, UK \\
	\And
	{\hspace{1mm}Gianluca Baio} \\
	Department of Statistical Science\\
	University College London\\
	London, UK \\
}

\date{}

\renewcommand{\headeright}{}
\renewcommand{\undertitle}{}
\renewcommand{\shorttitle}{}

\hypersetup{
pdftitle={A Bayesian hierarchical model for improving exercise rehabilitation in the ICU},
pdfsubject={},
pdfauthor={},
pdfkeywords={},
}

\begin{document}
\maketitle

\begin{abstract}
	Patients who are mechanically ventilated in the intensive care unit (ICU)  participate in  exercise as a component of their rehabilitation to ameliorate the long-term impact of critical illness on their physical function. The effective implementation of these programmes is hindered, however, by the lack of a scientific method for quantifying an individual patient's exercise intensity level in real time, which results in a broad one-size-fits-all approach to rehabilitation and sub-optimal patient outcomes. In this work we have developed a Bayesian hierarchical model with temporally correlated latent Gaussian processes to predict \(\dot VO_2\), a physiological measure of exercise intensity, using readily available physiological data. Inference was performed using Integrated Nested Laplace Approximation. For practical use by clinicians \(\dot VO_2\) was classified into exercise intensity categories. Internal validation using leave-one-patient-out cross-validation was conducted based on these classifications, and the role of probabilistic statements describing the classification uncertainty was investigated.
\end{abstract}

\keywords{Bayesian Hierarchical Modelling, Gaussian Processes, INLA, Exercise Rehabilitation, Critical Illness}

\section{Introduction}
\label{sec:Intro}

Patients who are mechanically ventilated in the intensive care unit (ICU) as a result of critical illness are often left with a range of impairments,  due to the pathological effects of critical illness and its treatments on nerve, muscle cardiac and respiratory function.\citep{Guarneri2008} This phenomenon is known as post-intensive care syndrome \citep{Myers20161} and can include a decline in physical, psychological and/or cognitive status. More than 224,000 patients are admitted to intensive care in the UK every year\citep{ICNARCReport2018} and only 50\% of those previously employed will have returned to work one year after admission due to ongoing health issues.\citep{Kamdar:2020aa} Reducing the impact of post-intensive care syndrome is therefore a crucial issue.

Rehabilitation, while the patient is still receiving mechanical ventilation in the ICU, involves progressing patients through activities such as sitting over the edge of the bed and standing and walking, and is considered the best way to ameliorate the impact of critical illness and its associated treatments on physical function. 

The current approach to ICU rehabilitation is based on the default assumption that the metabolic cost of individual rehabilitation activities does not differ across patients and is similar to that of healthy individuals, resulting in patients receiving a broad one-size-fits-all approach to rehabilitation. This assumption, however, was shown to be invalid by \cite{Black2020}, who found significant variation in absolute exercise intensity both between patients and within patients over time. This suggests that under the current approach to rehabilitation individuals with very different physiological profiles receive similar, and often sub-optimal, exercise programmes. 

Consequently, patients are often being under- or over-exercised, in the former case receiving minimal benefit from rehabilitation and in the latter being subjected to severe physiological stress. To address these issues, a scientific method for quantifying the exercise intensity level of an individual patient in real time is required.

Exercise load during rehabilitation can be quantified by measuring a patient's rate of oxygen consumption (\(\dot VO_2 ~ mL/kg.\textsuperscript{-1}min\textsuperscript{-1}\)), but these measurements are not available in many intensive care units. In addition, the equipment required to measure \(\dot VO_2\) is costly and requires technical expertise, meaning it is unlikely to be introduced as a regular feature in critical care in the foreseeable future.\citep{Black2015}

While \(\dot VO_2\) cannot be easily measured in the ICU, data are collected on a number of other physiological covariates as standard, many of which have known relationships to \(\dot VO_2\). 

The primary contribution of this paper is the development and internal validation of a first of its kind prediction model of \(\dot VO_2\) for mechanically ventilated intensive care patients, through the development of a Bayesian hierarchical model with temporally correlated latent Gaussian processes. This model can provide clinicians with real-time predictions of patients' levels of absolute exercise intensity, allowing tailored exercise rehabilitation plans to be implemented.

Adopting a Bayesian framework for this model allows us to easily account for both the hierarchical and temporal structure of the data. Further, when used in practice, it may be more useful for clinicians that the model returns classifications of exercise intensity based on \(\dot VO_2\) values, rather than predicted \(\dot VO_2\) values directly. Future clinical work is needed, however, to definitively define these categories, and different clinical scenarios may in practice require different classifications. Our approach allows us to adapt our model to any future categorisation required.

The structure of this paper is as follows. We begin with a presentation of the data (Section \ref{sec:BandEDA}), followed by a discussion of the model development process (Section \ref{sec:ModDev}) and a brief description of the INLA methodology used to perform inference (Section \ref{sec:INLA}). We present our results (Section \ref{sec:Results}) before assessing predictive performance (Section \ref{sec:Error}). We conclude with a discussion (Section \ref{sec:Disc}).

\section{Background and data}
\label{sec:BandEDA}

Rehabilitation in Intensive Care is considered fundamental to improve physical function outcomes for patients  post-ICU. However, an inability to easily quantify the exercise load leads to patients being over- or under-exerted.  Providing a reliable and accurate method for predicting \(\dot VO_2\) would allow exercise regimens to be individually tailored, potentially improving patient outcomes post-ICU.

Outside of the critical care environment, oxygen consumption  expressed in terms of \(\dot VO_2\), is the standard measure of quantifying the intensity of exercise. This is measured in millilitres per minute and is standardised by dividing by an individual's body weight.  To measure \(\dot VO_2\) breath by breath gas exchange analysis (BBGEA) is used, a common technique for healthy individuals but one that has only recently been applied to the critical care environment.\citep{Black2015,Som2019} While providing crucial information about the effect of exercise in the mechanically ventilated patients, this method requires a high level of technical expertise and is difficult to implement consistently.\citep{Black2020} It is therefore unlikely to become standard in intensive care in the near future. 

During rehabilitation, patients progress through a number of activities, including sitting up in bed, sitting over the edge of the bed and standing up. One possible modelling approach is to assess the impact of each of these activities on oxygen consumption for an individual. The \(\dot VO_2\) required for each of these activities, however, has been shown not just to vary between patients but also within patients over time \citep{Black2020} and so this alone would not be sufficient to capture exercise intensity.

Once measured, a patient's level of absolute exercise intensity could be summarised as being at rest (\(\dot VO_2 < 3.5\)), low (\(3.5 \leq \dot VO_2 < 5\)), medium (\(5 \leq \dot VO_2 < 7.5\)) or high (\(\dot VO_2 \geq 7.5\)), information which would allow clinicians to tailor rehabilitation activities in real time, with rest corresponding to a patient experiencing no exercise load, and high corresponding to a patient being excessively exerted. 

While these are arbitrary values to an extent they serve to illustrate the extremely limited exercise capacity of this patient group. For example; low level activity would equate to a healthy individual sitting over the edge of the bed, medium level of activity;  standing from a seated position and high level marching on the spot or transferring from a bed to chair.\citep{Ain2000}

\subsection{Data}
\label{sec:Data}

The data are from the observational study conducted by \cite{Black2020}, which used BBGEA to take measurements as mechanically ventilated ICU patients participated in various rehabilitation activities. All patients in the study had been persistently critically ill and or mechanically ventilated for more than 7 days. BBGEA records measurements on a breath by breath basis, resulting in high frequency data consisting of 74,332 measurements from 37 patients and 103 rehabilitation sessions. They are naturally hierarchical in nature with repeated measurements within sessions and often multiple sessions per patient.

Available physiological covariates are defined in Table \ref{tab:Vars}. Additionally standard baseline characteristics of patients were measured - height, weight, sex, age and the patient's pre-admission physical activity level given by the General Practice Physical Activity Questionnaire (GPPAQ). The Sequential Organ Failure Assessment (SOFA) score, a measure of organ system performance, was re-assessed before each session.

Finally the time since the start of the session and days since admission were also recorded, as well as the quality of session as assessed by the clinician running the study (classified as good, reasonable or poor), with poorer quality sessions being potentially more susceptible to measurement error in any of the physiological covariates, typically due to instability of the ventilator delivered oxygen levels and patients coughing.

\begin{table}
\bgroup
\def\arraystretch{1.2}
\begin{tabular}{|p{2cm}|p{14cm}|}
\hline
\textbf{Variable}                        & \textbf{Description}         \\
\hline\hline
\(\dot VO_2\)                     &     Volume of oxygen uptake. Values are not readily available for most intensive care patients and are measured through BBGEA.    \\
\hline
\( V_T\)            &   Tidal volume, the amount of air moved in and out of the lungs during a single breath.           \\
\hline
\(RR\)       &      Respiratory rate, the number of breaths per minute.          \\
\hline
\(\dot V_E\)          &        Minute ventilation, the amount of air breathed per minute, defined as the product of tidal volume and respiratory rate.      \\
\hline
\(P_{ET}CO_2\)                &      Amount of CO\textsubscript{2} in exhaled air on a breath-by-breath basis.            \\
\hline
\end{tabular}
\caption{Physiological variables in the data set.}
\label{tab:Vars}
\egroup
\end{table}

\subsubsection{Exploratory analysis}
\label{sec:EDA}

\begin{figure}[!h]
\includegraphics[width = 16cm]{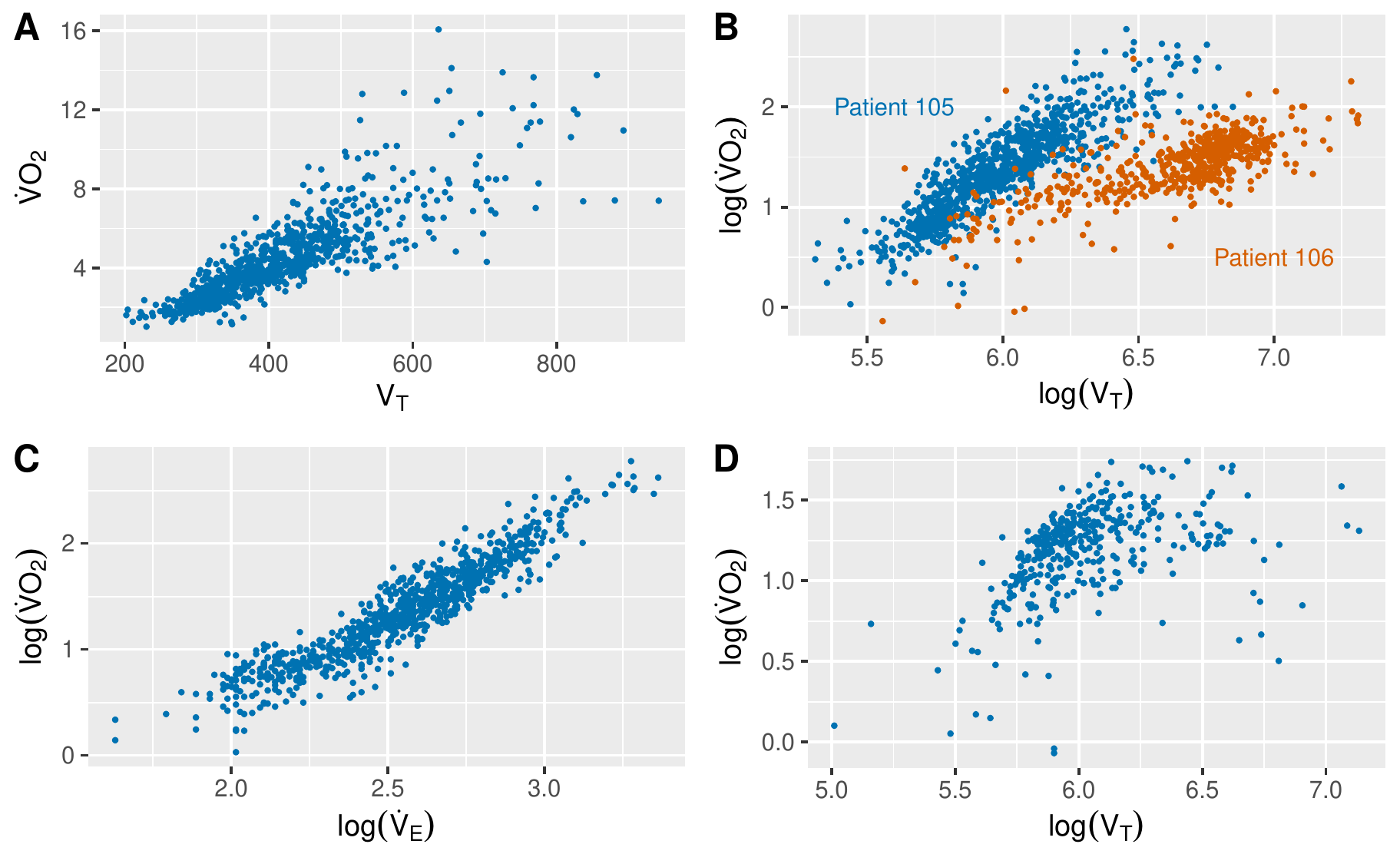}
\caption{Relationships between \(\dot VO_2\) and the physiological covariates in the data. (A) \(\dot VO_2\) against \(V_T\). (B) \(\dot VO_2\) against \(V_T\) on the log-log scale for patients 105 and 106 to show the heterogeneity in the effect of \(V_T\) on \(\dot VO_2\). (C) \(\dot VO_2\) against \( \dot V_E\) on the log-log scale. (D) \(\dot VO_2\) against \(V_T\) on the log-log scale for patient 117, where the relationship is decidedly non-linear. }
\label{fig:2ScatterPlots}
\end{figure}

Exploratory analysis revealed a number of relationships between covariates in the data. Figure \ref{fig:2ScatterPlots} features example plots of these relationships for individual rehabilitation sessions. Figure \ref{fig:2ScatterPlots}A shows us that while there is a non-linear relationship between \(V_T\) and \(\dot VO_2\) on the natural scale, this relationship is brought closer to linearity when placed on the log-log scale (see Figure \ref{fig:2ScatterPlots}B). Figure \ref{fig:2ScatterPlots}C shows us that this relationship similarly holds for \( \dot V_E\) - unsurprisingly given that \(\dot V_E\) is the product of \(V_T\) and respiratory rate.

Figure \ref{fig:2ScatterPlots}B features \(V_T\) and \(\dot VO_2\) measurements from two patients. Note how for each patient, the relationship between \(V_T\) and \(\dot VO_2\) appears linear, but the gradient differs, indicating that there is between-patient heterogeneity present. Furthermore, for a small minority of sessions, these relationships were not linear even on the log-log scale as shown in Figure \ref{fig:2ScatterPlots}D. 

For both respiratory rate and \(P_{ET}CO_2\) there was weak evidence for linear relationships with \(\dot VO_2\) when placed on the log scale, however examining plots of their product with \(\log(V_T)\) against \(\dot VO_2\) suggested that interactions may be present. Finally an interaction between age and BMI was also identified graphically, as shown in the supporting information.

\begin{figure}[h]
\includegraphics[width = 16cm]{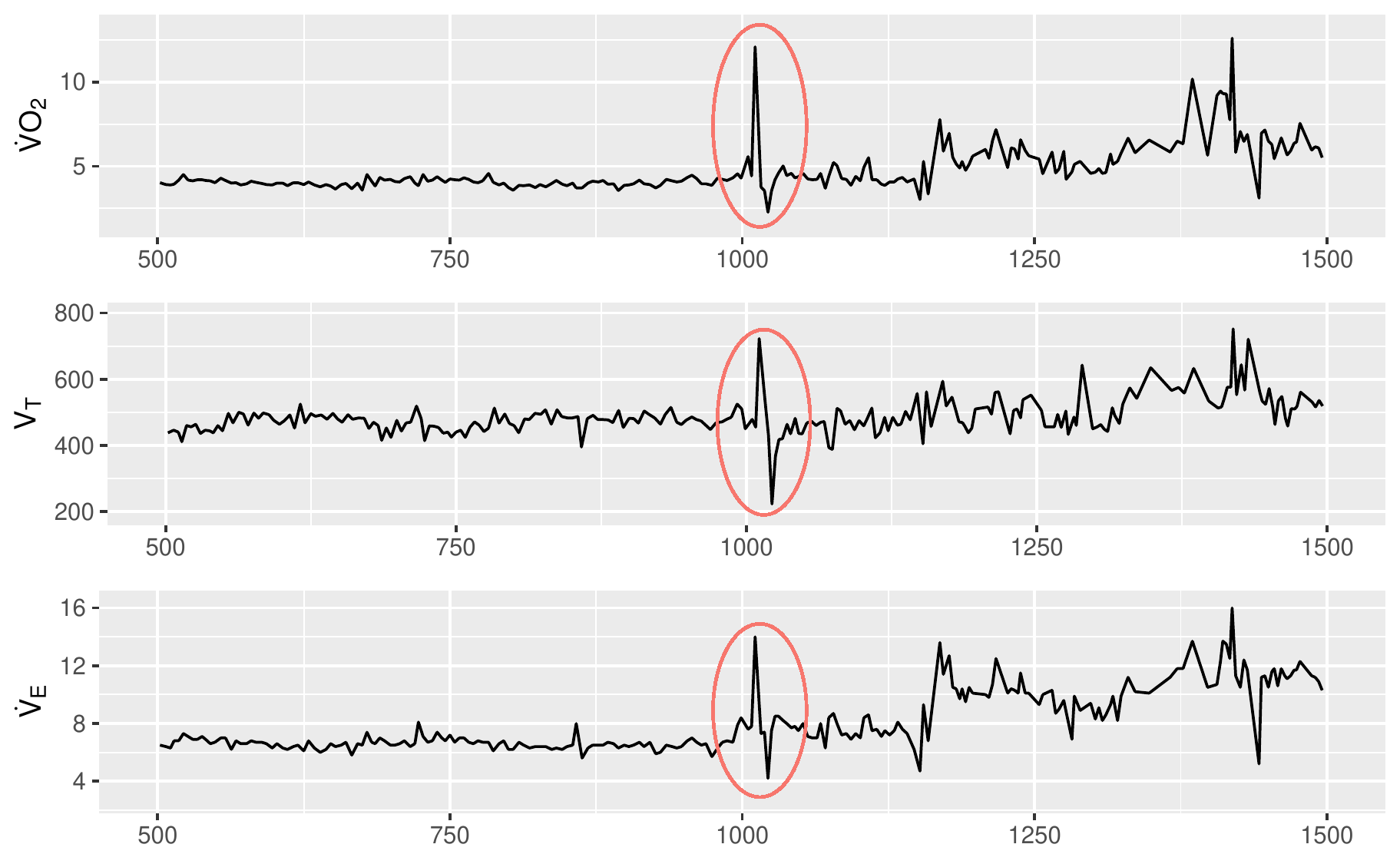}
\caption{Plots showing \(\dot VO_2\), \(\dot V_E\) and \( V_T\) over time for session 73, with disruption caused by a cough circled.}
\label{fig:2OverTime}
\end{figure}

\subsubsection{Data features}
\label{sec:DataFeatures}

The data have two notable features worth mentioning here. Firstly, patients were liable to cough during rehabilitation sessions, potentially disrupting the relationships between covariates. The effect of coughing on \(\dot VO_2\), \( V_T\) and \(\dot V_E\) is circled in Figure \ref{fig:2OverTime}. We further assess the impact of coughing and smoothing out coughs in Section \ref{sec:Results}.

Secondly, measurements were taken over time and on a breath by breath basis. The result is that as respiratory rate increases the rate of measurements increases and vice versa, resulting in inconsistent intervals between measurements.

The data were subject to measurement error from multiple sources. The BBGEA methodology is imperfect and may naturally lead to incorrect measurements for all the physiological covariates. The most obviously incorrect values these produced were removed from the data using set criteria, but incorrect values will still be present.

Finally, the patient population was relatively homogeneous in terms of age and BMI, however, we removed two patients from the analysis aged < 40 who were significantly younger than the rest of the population, and exhibited different physiological behaviour, in terms of the relationship between \(V_T\) and \(\dot VO_2\). The minimum, lower quartile, median, upper quartile and maximum values for age after their removal was \((50.2, 58.0, 69.7, 77.4, 86.0)\).

\subsection{Existing physiological relationships}
\label{sec:PhysRels}

Documented relationships already exist between \(\dot VO_2\) and the physiological measures introduced in the preceding sections. These have helped motivate this work and so for completeness we provide a brief overview of them here. As these relationships only hold in specific instances,\citep{Ward2007} however, they were not used in developing the model. 

The relationship that has received the most attention in previous research is that between \(\dot VO_2\) and minute ventilation (\(\dot V_E\)).\citep{Ramos2013}  An existing relationship between the two quantities is the Oxygen Uptake Efficiency Slope (OUES) \citep{Baba1996}, where oxygen uptake efficiency is the amount of gas an individual needs to breathe in and out in order to utilise a given volume of oxygen. 

The OUES is defined by the equation \(\dot VO_2 = a \times \log(\dot V_E) + b\), and is simply the change in \(\dot VO_2\) as \(\log(\dot V_E)\) increases.  The OUES is significant in the exercise intensity literature as a measure of physical fitness independent of a patient's motivation to exercise.\citep{Baba1996} This can also be extended to further physiological covariates as minute ventilation is defined as the product of tidal volume (\( V_T\)) and respiratory rate (\(RR\)).

A second way of connecting \(\dot V_E\) and \(\dot VO_2\) is through the relationship between \(\dot V_E\) and \(VCO_2\) - the volume of exhaled carbon dioxide. Essentially \(\dot V_E\) can be written as a function of \(VCO_2\) and other physiological variables including \(V_T\) and \(P_{ET}CO_2\), which can then be linked to \(\dot VO_2\) as the ratio of \(\dot VO_2\) and \(VCO_2\) is known to take values between 0.7 and 1.2. We provide further information on these covariates in the supporting information, as noted above however, as these relationships do not hold universally we will not be utlising them further in developing the prediction model.

\section{Model development}
\label{sec:ModDev}

We will be undertaking model development within the Bayesian paradigm. In practice predictions of \(\dot VO_2\) would likely be presented to clinicians in the form of classifications; the Bayesian approach allows for uncertainty around \(\dot VO_2\) predictions to be propagated through the model, and used to quantify uncertainty about classifications. Future work is needed to determine these classification categories, however, and in practice different categories may be used in different contexts. This modelling approach allows us to adapt the model to any classification task that may arise.

\subsection{The model}
\label{sec:TheMod}

In order to account for the structured nature of the data (see Section \ref{sec:Data}) we have developed a Bayesian hierarchical model for \(\dot VO_2\) with temporally correlated latent Gaussian processes. A graphical representation of the modelling assumptions is presented in the supporting information, in the form of a directed acyclic graph. 


We indicate the value of \(\log (\dot VO_2)\) taken at time \(t\) for the \(i^{th}\) patient's \(j^{th}\) rehabilitation session as \(y_{ijt}\) and assume that
\begin{equation}
\label{eq:VO2Normal}
    y_{ijt} \mid \mu_{ijt}, \tau \sim \text{Normal}(\mu_{ijt}, \tau^{-1}),
\end{equation}
where $\tau=\sigma^{-2}$ indicates the precision.

We then specify \(\mu_{ijt}\) as a linear predictor of the form
\begin{align}
\nonumber
    \mu_{ijt} =\quad &\alpha_{ij} + \beta_{1i}\log(V_{T_{ijt}}) + \beta_{2}\log(P_{ET}CO_{2_{ijt}}) + \beta_{3}\log(RR_{ijt})\ +\\
\label{eqn:Model}
     &\beta_{4}\log( V_{T_{ijt}}) \times \log(P_{ET}CO_{2ijt}) + \beta_{5}\log( V_{T_{ijt}}) \times \log(RR_{ijt})\ + \\
\nonumber
     &[\ldots] + s_{ijt} \ .
\end{align}
Note that in the above [\ldots] is used to indicate adjustment for session and patient level covariates --- in this case age, BMI, GPPAQ and SOFA score, as well as an interaction between age and BMI as identified in Section \ref{sec:EDA}. Covariates were included based on empirical findings (Section \ref{sec:EDA}) and their clincial relevance as determined by a clinician. Although, as shown in Figure \ref{fig:2ScatterPlots} (C), \(\dot V_E\) has a strong linear relationship with \(\dot VO_2\), we have chosen to decompose \( \dot V_E\) into \(V_T\) and respiratory rate to maximise the information provided to our model.

All continuous covariates and \(\dot VO_2\) are placed on the log-scale and centered around their sample means to aid interpretation and for numerical stability. Further, pairwise interactions between \(V_T\) and both \(P_{ET}CO_2\) and respiratory rate are included as motivated by the findings in Section \ref{sec:EDA}. The unstructured coefficients were given relatively ``minimally informative'' Normal\((0,0.1^{-1})\) priors, corresponding to a prior belief that seeing effect sizes of over 6.3 was highly unlikely. The specification of the precision \(\tau\) is discussed in Section \ref{sec:Temporal}.

In Section \ref{sec:Data} we noted both the hierarchical and temporal nature of the data. The remainder of this section is focused on how the model accounts for this structure.

\subsubsection{Structured effects}
\label{sec:Hierarchical}

The between-patient and between-session heterogeneity in the model is accounted for through two terms in \eqref{eqn:Model} --- a session-level varying intercept term and a patient-level varying coefficient for the effect of \(\log(V_T)\). These are defined as
\begin{align}
    \nonumber
    \alpha_{ij} \mid \mu_\alpha, \tau_{\alpha} \sim \text{Normal}(\mu_\alpha, \tau_{\alpha}^{-1}), \\
    \nonumber
    \beta_{1i} \mid \mu_{\beta_1}, \tau_{\beta_1} \sim \text{Normal}(\mu_{\beta_1}, \tau_{\beta_1}^{-1}),
\end{align}
with Normal\((0,0.1^{-1})\) priors for \(\mu_{\alpha}\) and \(\mu_{\beta_1}\) and vague, proper log-Gamma priors for \(\tau_\alpha\) and \(\tau_{\beta_1}\).

\begin{figure}
\includegraphics[width = 16cm]{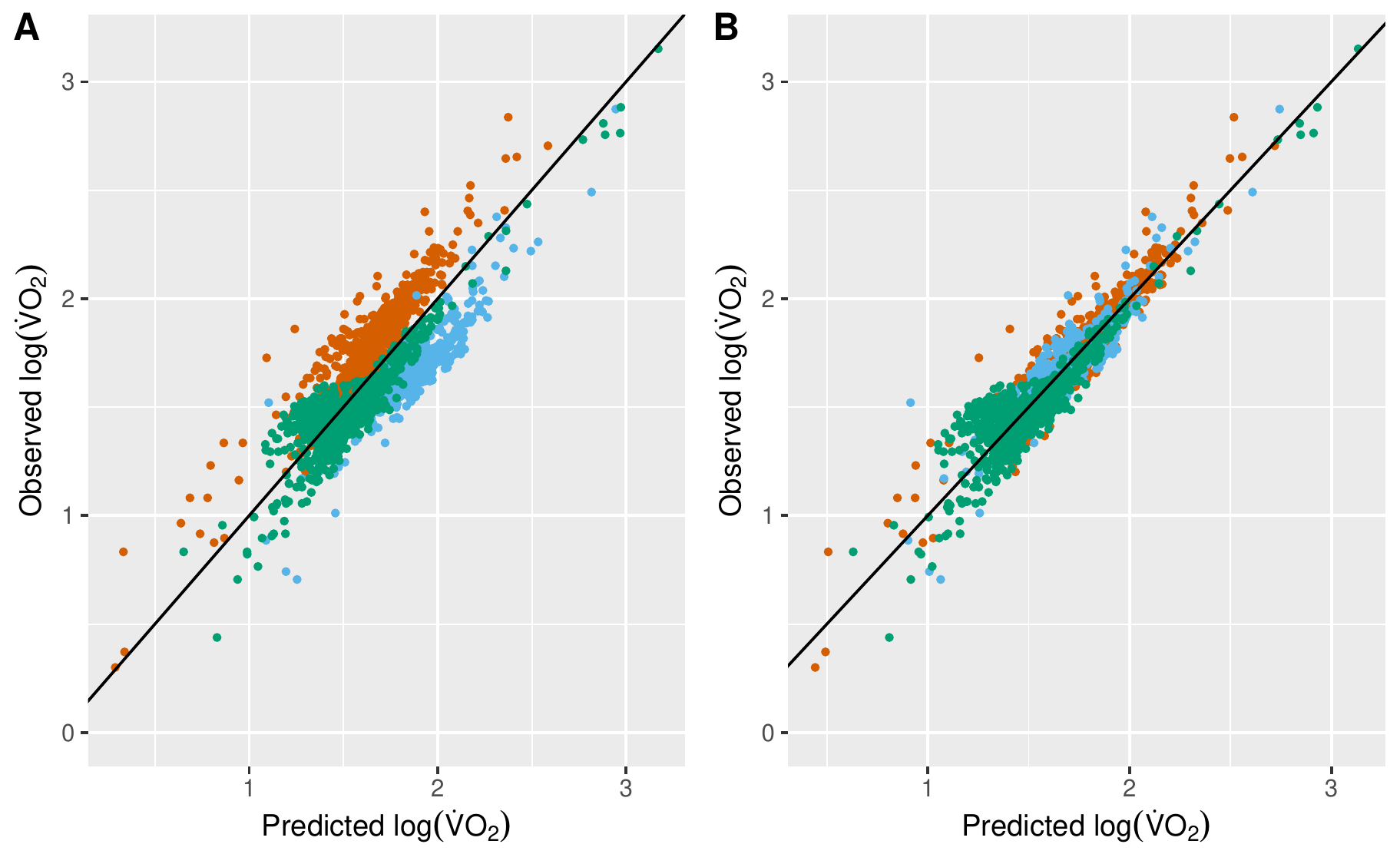}
\caption{Plots of predicted \(\dot VO_2\) against actual \(\dot VO_2\) for patient 112, coloured by session for models with individual intercepts at the patient (A) and session (B) level. There is clear between session heterogeneity in (A) which is only resolved with the session level intercept in (B). }
\label{fig:4intercepts}
\end{figure}

Plots of predicted \(\log(\dot VO_2)\) values against observed \(\log(\dot VO_2)\) for these two models can be seen in Figure \ref{fig:4intercepts} with points highlighted by session. Even after accounting for between patient heterogeneity, from Figure \ref{fig:4intercepts} we can see that there is still some between session variation. This was captured through a hierarchical prior structure at the session level.

We noted in Figure \ref{fig:2ScatterPlots} evidence for a heterogeneous relationship between \(V_T\) and \(\dot VO_2\), which motivates the inclusion of individual patient-level coefficients. To examine further whether individual coefficients were required, we re-fit the model separately for each patient and then examined the resulting posteriors for the \(V_T\) coefficient. The effect of \(V_T\) on \(\dot VO_2\) had a high level of heterogeneity between patients. Although this is likely to be an over-estimate of the heterogeneity in the data, this does indicate we need to account for a heterogeneous effect of \(V_T\) at the patient level. Graphical summaries of these posteriors can be found in the supporting information.

\subsubsection{Temporal effects}
\label{sec:Temporal}

The \(s_{ijt}\) term in \eqref{eqn:Model} is a temporal error term. We model this using an Ornstein-Uhlenbeck process (O-U process).\cite{Oksendal2003} This is a Markov process such that if the previous realisation of the process was at time \(r < t\), we have
\begin{equation*}
    s_{ijt} \mid \phi, \tau_s, s_{ijr} \sim \text{Normal}(\mu_{*},\tau_{*}^{-1}), \\
\end{equation*}
with 
\begin{align*}
    \mu_{*} &= s_{ijr}\exp(-\phi\vert t-r\vert) , \\
    \tau_{*} &= \tau_s\Big(1 - \exp\left(-2\phi\vert t-r\vert\right)\Big)^{-1} .\\
\end{align*}
Further, if \(t\) is the first observation time for the \(j^{th}\) session, then
\begin{equation*}
    s_{ijt} \sim \text{Normal}(0,\tau_s^{-1}) .
\end{equation*}
Here \(\tau_s\) is the precision of the process at stationarity and \(\phi\) is a mean-reversion parameter, which intuitively represents the level of similarity between consecutive observations, with smaller values of \(\phi\) indicating higher levels of similarity. While a vague Normal prior was used for \(\log(\phi)\), a weakly informative log-gamma(50,1) prior for \(\tau_s\) was used to stabilise the inference.

Temporal dependence between observations is more commonly characterised using the discrete-time analogue to the O-U process --- the \(AR(1)\) process.\citep{Blangiardo2013} The primary advantage of using the O-U process in this setting is that it naturally accounts for irregular intervals between observations, which arise from breath by breath measurements. 

Finally, in the current model specification we have two sources of measurement level variance: \(\tau\) in \eqref{eq:VO2Normal} and the O-U process, potentially leading to issues of non-identifiability. To address this we fix \(\tau = e^{15}\) as seen in \cite{Wang2018}, in effect forcing all the observational error to appear through the temporal error term. As well as ensuring identifiability of the model this has the added benefit of reducing complexity, by removing one hyperparameter from the posterior that needs to be computed. This is particularly computationally advantageous, especially in the context of our approach, as described in Section \ref{sec:INLA}. 

We performed sensitivity analysis to assess the impact of the prior choices made in this section on our model. Separate models were fitted replacing the O-U process prior with a first order random walk prior and replacing fixed \(\tau = e^{15}\), with a strongly informative prior for \(\tau\) centered around 1000. In both cases neither change in prior resulted in a significant change of direction or magnitude in the marginal posteriors of the model coefficients.

\subsection{Inference}
\label{sec:INLA}

There are two primary inferential challenges presented by this model. Firstly, the varying intercept and \(V_T\) coefficient terms naturally introduce non-linearity skewing the posterior and secondly, due to the presence of temporal effects, the dimension of the posterior grows with the number of observations, and there is often (by construction) significant posterior dependence between the high-dimensional latent variable and the hyperparameters associated with it. Both of these challenges can significantly hamper the performance of traditional MCMC methods. The model presented in \eqref{eqn:Model} was therefore fitted using Integrated Nested Laplace Approximation (INLA), with the \texttt{R-INLA} package.\citep{Rue2017} 

The principal idea underlying INLA is to focus on approximating the marginal posterior distributions of latent Gaussian models (LGMs) using Laplace approximations and numerical integration. The most relevant subclass of LGMs for this work are based on the linear predictor 
\begin{equation}
\label{eq:LGM_linpred}
\eta_i = \beta_0 + \sum_{j = 1}^J\beta_jx_{ij} + \sum_{k=1}^Kf_k(z_{ik}) + \epsilon_i \;, \\
\end{equation}
which is linked to observations \(\mathbf{y}\) from an exponential family via a link function and where \(\mathbf{\theta} = (\mathbf{\beta}, \mathbf{f})\) is given a Gaussian prior conditional on \(\mathbf{\psi}\), a set of hyperparameters. Here \(\mathbf{x}\) is a matrix of covariates and \(\mathbf{f} = \{f_k\}_{k = 1}^K\) is a set of functions dependent on covariates \(\mathbf{z}\) which, conditionally on the choice of function, can be used to define hierarchical, spatial and temporal effects. Note that the model in Section \ref{sec:TheMod} naturally falls into this definition of an LGM.

INLA numerically approximates the marginals of interest by first computing the posteriors of \(\mathbf{\theta}\) and the conditional \(\mathbf{\psi}\) using repeated nested Laplace approximations. The mode of the posterior is then found using numerical optimisation and marginal posteriors are computed using numerical intergration over \(\mathbf{\psi}\). Comprehensive reviews of the method applied in different contexts are provided in \cite{Rue2017},\cite{Blangiardo2013} and \cite{Wang2018}.

The primary advantage of using INLA over simulation-based methods, is that the cheap cost of the Laplace approximations means that \(\dim(\mathbf{\theta})\) can grow rapidly with only minor impact on computation time. The numerical integration of $\mathbf{\psi}$ means that complex dependencies between this and $\mathbf{\theta}$ can be dealt with effectively, albeit at the cost of enforcing that the dimension of $\mathbf{\psi}$ is low. This is particularly useful for the model in Section \ref{sec:TheMod}.

\section{Results}
\label{sec:Results}

Table \ref{tab:4Posts} shows posterior summaries for both the model hyperparameters and the fixed effects. The posteriors for the coefficients of the physiological covariates indicate that the effect of each covariate on \(\dot VO_2\) is positive, as expected following Section \ref{sec:EDA}. Further, given the posteriors for the interactions, the effect of \(\log( V_T)\) increases as respiratory rate increases and decreases as \(P_{ET}CO_2\) increases. Examining the session and patient level covariates, their effects are either not significant or are in the expected direction.

For the hyperparameters, the posterior summaries of \(\tau_\alpha\) and \(\tau_\beta\) both indicate that there is non-negligible heterogeneity in the model and that their inclusion is therefore necessary. Examining the temporal effect, the posterior for \(\phi\) has low uncertainty and is focused around 0.09, indicating a high level of similarity between consecutive observations. 

\begin{center}
\begin{table}[!ht]%
\centering
\caption{Posterior summaries for the fixed effects coefficients and hyperparameters of the model defined in \eqref{eqn:Model}, with mean, standard deviation (SD) and a 95\% central posterior interval. ($\dagger$ Fixed effects with 95\% posterior credible intervals that do not contain 0).\label{tab:4Posts}}%
\begin{tabular*}{350pt}{@{\extracolsep\fill}lrrrr@{\extracolsep\fill}}
\toprule
 & \textbf{Mean}  & \textbf{SD}  & \textbf{2.5\%}  & \textbf{97.5\%} \\
\midrule
Intercept\(\dagger\) & 1.32 & 0.09 & 1.14  & 1.50 \\ 
  \(\log ( V_T)\dagger\) & 1.56 & 0.03 & 1.52  & 1.61 \\ 
  \(\log (P_{ET}CO_2)\dagger\) & 1.92 & 0.01 & 1.91  & 1.93  \\ 
  \(\log (RR)\dagger\) & 1.11 & \(<\)0.01 & 1.10  & 1.11   \\ 
  \(\log ( V_T) \times \log (P_{ET}CO_2)\dagger\) & -0.20 & 0.01 & -0.22  & -0.17 \\ 
  \(\log ( V_T) \times \log (RR)\dagger\) & 0.33 & \(<\)0.01 & 0.32  & 0.34   \\ 
  SOFA & -0.01 & 0.02 & -0.05  & 0.03   \\ 
  GPPAQ = 2 & -0.01 & 0.05 & -0.11  & 0.09 \\ 
  GPPAQ = 3 & 0.01 & 0.05 & -0.09  & 0.11   \\ 
  GPPAQ = 4\(\dagger\) & 0.31 & 0.08 & 0.16  & 0.46  \\ 
  Sex\(\dagger\) & -0.08 & 0.04 & -0.17 & 0.00  \\ 
  \(\log(age)\dagger\) & 0.35 & 0.12 & 0.11  & 0.59  \\ 
  \(\log(BMI)\dagger\) & -1.13 & 0.09 & -1.31  & -0.96   \\ 
  \(\log(age) \times \log(BMI)\dagger\) & -2.12 & 0.65 & -3.39  & -0.96  \\ 
  \midrule
  \(\tau_\alpha\) & 38.63 & 3.56 & 31.00  & 44.65  \\ 
  \(\tau_\beta\) & 44.30 & 6.27 & 31.54  & 55.82  \\ 
  O-U \(\tau\) & 46.75 & 0.71 & 45.17  & 47.90  \\ 
  O-U \(\phi\) & 0.09 & \(<\)0.01 & 0.09  & 0.10 \\ 
\bottomrule
\end{tabular*}

\end{table}
\end{center}
\subsection{Smoothing}
\label{sec:Smoothing}
In Section \ref{sec:BandEDA} we noted that patients coughing may induce spikes in the data corresponding to periods of time we are not looking to directly model. To investigate the impact of these coughs we applied a ``smoothing'' three-value rolling average to the physiological covariates in the data and re-fit the model. An example of the effect of smoothing can be seen in Figure \ref{fig:4PostPredCheck}. Five- and seven- value rolling averages were also tested, but the resulting loss of resolution in the data was deemed too high to be able to accurately model cases of high exercise intensity.

Selected posteriors of the raw and smoothed data models are plotted in Figure \ref{fig:4PostRobust} and full posterior summaries are available in the supporting information. Naturally there is shrinkage towards zero in the posterior estimates, but the direction and magnitude of all effects remains unchanged. These results suggest that the model is not unduly effected by spikes in the data. A version of the model using smoothed covariates may have other benefits, however, which we explore further in Section \ref{sec:Error}.

\subsection{Robustness checks}
\label{sec:qual}
To ensure the robustness of the model in light of potential measurement error, we re-fit it to a subset of the data excluding poor quality sessions and another smaller subset excluding both poor and reasonable quality sessions. Full posterior summaries for these can be seen in the supporting information.

\begin{figure}[!ht]
\includegraphics[width = 16cm]{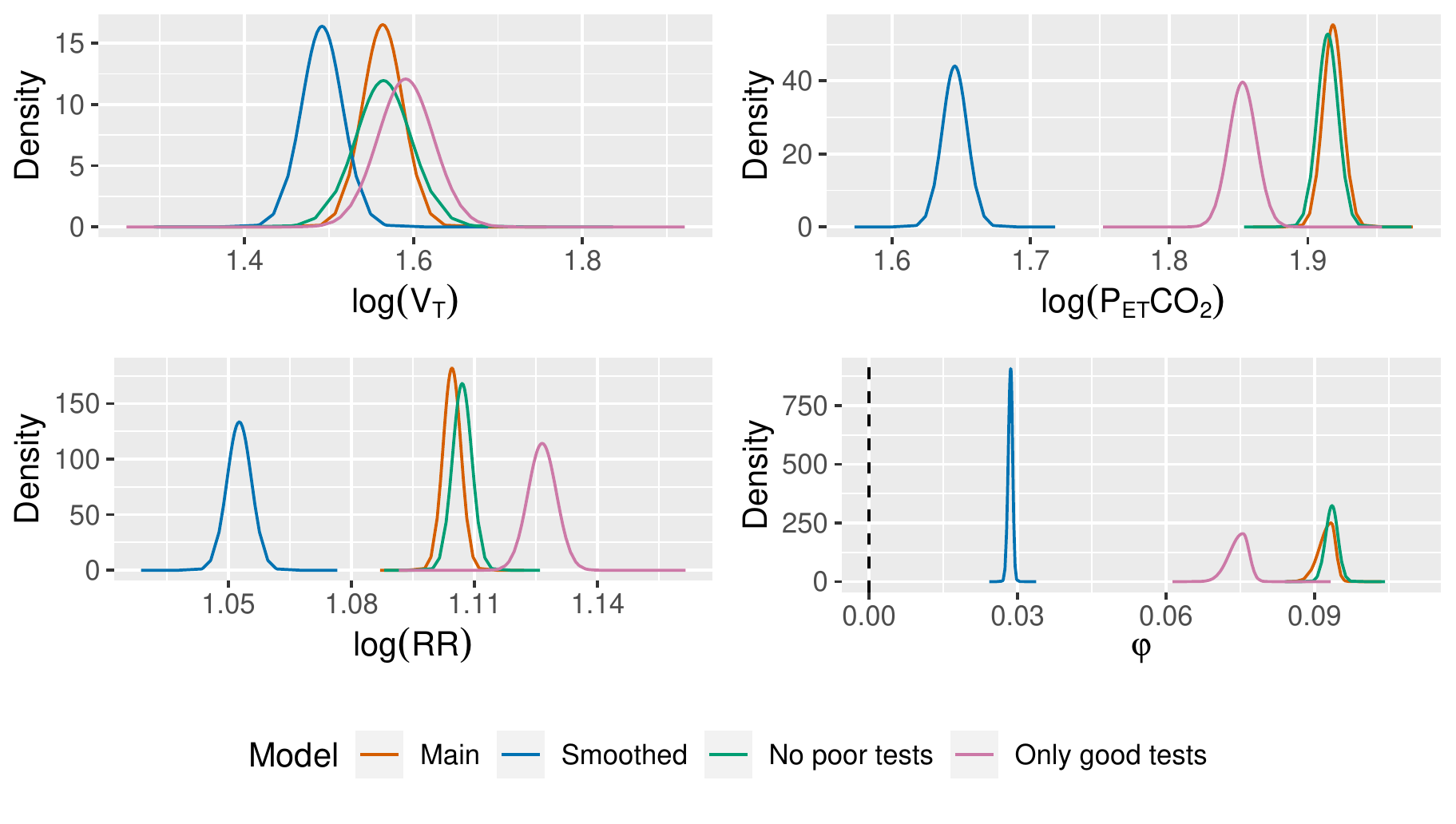}
\caption{Marginal posteriors for the  coefficients of \(\log( V_T)\), \(\log(P_{ET}CO_2)\), \(\log(RR)\) and the \(\phi\) hyperparameter of the O-U process, for the model in \eqref{eqn:Model}, the model fit using smoothed covariates, the model fit excluding poor quality sessions and the model fit excluding both poor and reasonable quality sessions. In the plot for \(\phi\), the dashed line is used to indicate 0.}
\label{fig:4PostRobust}
\end{figure}

The resulting posteriors for \( V_T\), \(P_{ET}CO_2\), respiratory rate and \(\phi\) are shown in Figure \ref{fig:4PostRobust}. There is a natural increase in the variance of the posteriors as we have a smaller sample size and a slight shift in their centres. Importantly, however, there is no major shift in their modes or in their direction, indicating the model is robust and has not been unduly influenced by potentially poorly measured values.

\section{Validation}
\label{sec:Error}

The gold standard for validating a model's predictive ability is external validation. However, as noted in Section \ref{sec:BandEDA}, the novel nature of the data means that this is not possible, in our case. We therefore used internal validation to assess the out of sample error of the model. This is done first through in time posterior predictive checks, to understand how the model performs given maximal information, and then secondly using leave-one-patient-out cross-validation to understand how the model performs in realistic conditions.

\subsection{Posterior predictive checks}
\label{sec:PostPredCheck}
To perform posterior predictive checks in light of hierarchical and temporal structure we re-fit the model as in Section \ref{sec:ModDev} except for data from session \(i\), where we only use data observed before time \(t = 1000\). We then use the model to predict \(\dot VO_2\) for \(t > 1000\) for session \(i\). Here \(t = 1000\) was chosen so that enough data were present to ensure model stability for the new patient and to allow for an initial calibration phase at the start of each session during which patient \(\dot VO_2\) values were largely in the ``rest'' category.

\begin{figure}[!ht]
\includegraphics[width = 16cm]{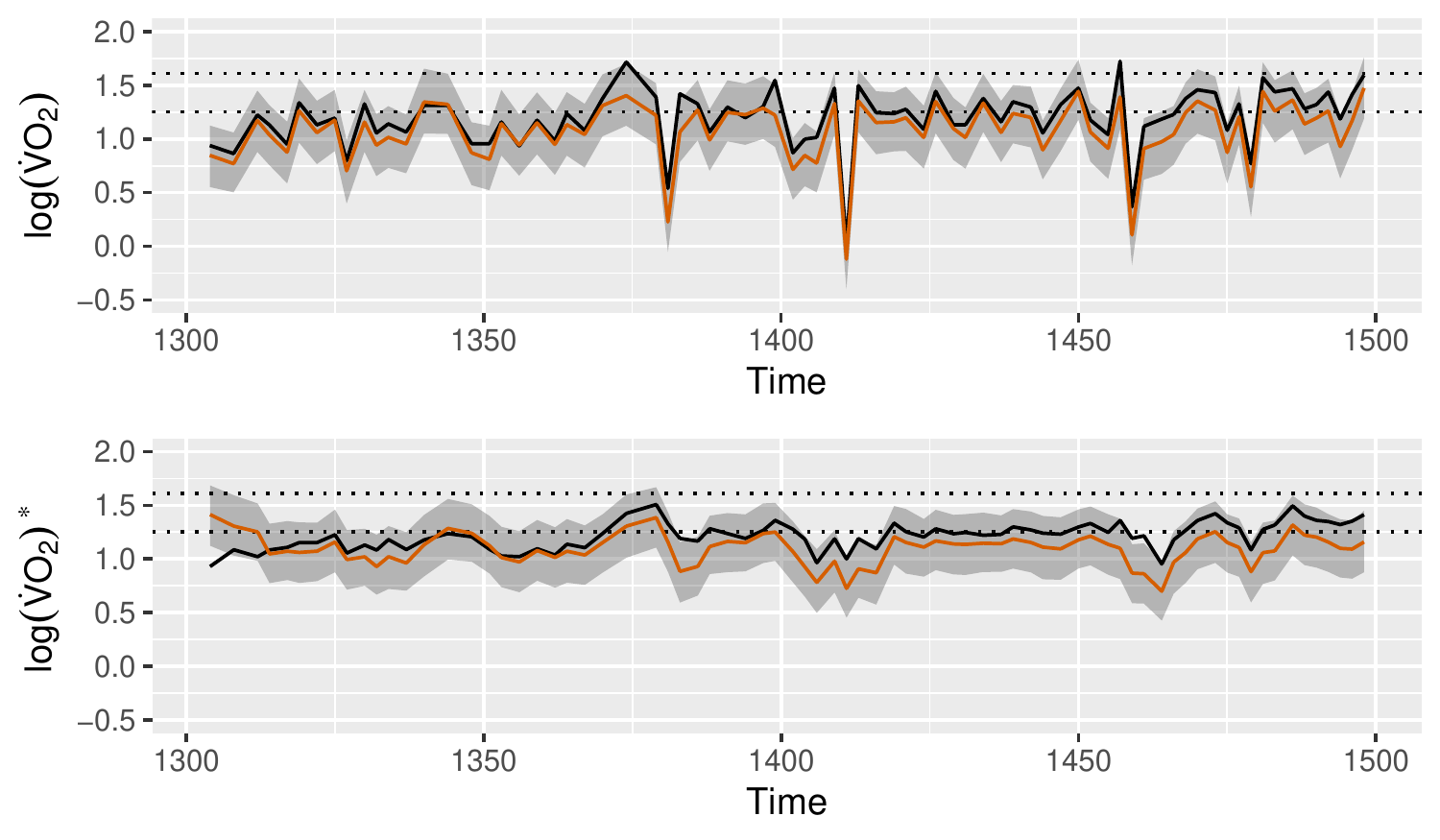}
\caption{An example of posterior predictive checks using cross-validation in time for patient 137, for the models using raw and smoothed data, as described in Section \ref{sec:PostPredCheck}. We use \(\log(\dot VO_2)^*\) to denote smoothed \(\dot VO_2\) values.}
\label{fig:4PostPredCheck}
\end{figure}

The results of these checks are shown in Figure \ref{fig:4PostPredCheck} for the raw and smoothed data models, with observed (black) and predicted (red) values of \(\log(\dot VO_2)\). The majority of model heterogeneity comes at the session and patient levels, meaning that credible intervals are relatively small here. We can see that the model does a remarkable job of matching the shape of \(\dot VO_2\), in both instances.

\subsection{Leave-one-patient-out cross-validation}
\label{sec:CV}
To conduct cross-validation observations were left out at the patient level, to account for similarity between observations within patients and prevent leakage. This resulted in 35 sets of observations to assess on re-fitted models. These were fit using the same specification as in Section \ref{sec:ModDev}.

Here, unlike in Section \ref{sec:PostPredCheck}, the model is unable to learn individual coefficients, increasing the uncertainty in the classifications. Given that clinicians would not have access to \(\dot VO_2\) values during regular rehabilitation sessions, this aligns with how the model would be used in practice.

To obtain the posterior predictive densities for each point, the \texttt{joint.posterior.sample} function in \texttt{R-INLA} was used to generate 1000 samples from the joint posterior, \(\pi^{(-i)}(\boldsymbol{x},\boldsymbol{\theta}\mid\mathbf{y})\), of the model fitted without patient \(i\). The samples were then fit to the data of the new patient, with uncertainty from the individual coefficients and temporal effects incorporated analytically.

Probabilities for each category were then easily generated using the proportion of samples from the posterior predictive density that fell into that category and, as this is the information that would be available to clinicians in practice, we classified the observations based on the category with the highest probability assigned to it.

\subsection{Quantifying predictive ability}
\label{sec:PredAbl}
To evaluate the predictive performance of the models after cross-validation we use two metrics. The first of these is the average zero-one loss, which is simply the proportion of observations incorrectly classified, with lower values indicating better predictive ability.\citep{Claude2010}

This is an intuitive measure and a proper scoring rule,\citep{Gnieting2004} however it does not take into account key information from the model. First, we have developed a model which, rather than simply classifying observations, also assigns probabilities to them. We would therefore like a method of assessment that also accounts for the confidence in the classification. Furthermore, the categories are not nominal but ordinal. Though incorrect, a prediction of low exercise intensity when the patient is at rest is therefore presumably better than a prediction of high and the measure should account for this.

Both these goals can be achieved using the ranked probability score \citep{Epstein1969} defined as follows. Given \(R\) mutually exclusive, ordinal outcomes, let \(p_1,...,p_R\) be the predicted probabilities associated with each outcome such that \(\sum_{i=1}^Rp_i = 1\) and let \(a_1,...,a_R\) be such that \(a_j = 1\) if the \(j\)-th outcome was observed and 0 otherwise. For a given measurement and predictions, the ranked probability score is then defined as 
\begin{equation*}
\mbox{RPS} = \frac{1}{r-1}\sum^{r-1}_{i=1}\left(\sum^i_{j = 1}(p_j - a_j)\right)^2.
\end{equation*}
This can then be summarised for a given model by averaging over the ranked probability score for all observations.

\subsection{Validation results}
\label{sec:Val_Results}
\begin{table*}[!h]
\centering
\caption{Confusion matrices for the predictive accuracy of the raw and smoothed data models as assessed using leave-one-patient-out cross-validation. \(\mathcal{L}\) = Average 0-1 loss, RPS = ranked probability score. }
\begin{tabular*}{350pt}{r@{\hskip 0.5cm}rrrr@{\hskip 1cm}rrrr}
\toprule
& &\multicolumn{2}{c}{\textbf{Raw data}} & & & \multicolumn{2}{c}{\textbf{Smoothed}} \\ 
\cmidrule{3-4}\cmidrule{7-8}
\textbf{Observed} & \textbf{Rest}  & \textbf{Low}  & \textbf{Med.}  & \textbf{High} & \textbf{Rest}
& \textbf{Low} & \textbf{Med.} & \textbf{High}\\
\midrule
Rest & 0.73  & 0.23  & 0.04  & \(<\)0.01 & 0.69 & 0.29 & 0.02 & \(<\)0.01   \\
Low & 0.36  & 0.43  & 0.21  & 0.01 & 0.33 & 0.50 & 0.16 & 0.01  \\
Medium & 0.09  & 0.26  & 0.47  & 0.18  & 0.06 & 0.30 & 0.52 & 0.12 \\
High & \(<\)0.01  & 0.03  & 0.22  & 0.74  & 0.01 & 0.07 & 0.36 & 0.56 \\
\midrule
& \multicolumn{2}{l}{\(\mathcal{L}\) = 0.400} & \multicolumn{2}{l}{RPS = 0.105 } &
\multicolumn{2}{l}{\(\mathcal{L}\) = 0.389} & \multicolumn{2}{l}{RPS = 0.101} 
\end{tabular*}
\label{tab:ConfMatrices}
\end{table*}
The results of this cross-validation procedure for both the raw and smoothed data models can be seen in Table \ref{tab:ConfMatrices}. In terms of both zero-one loss and ranked probability score, the models are very similar with the smoothed model performing marginally better by both metrics.

There is large variation in performance between observed exercise intensity categories, with both models having higher accuracy for resting and high observations. Here the open-ended nature of both of these categories means the model is not penalised for over-predicting high observations and under-predicting low observations.

These results also suggest that the model has a tendency to under-predict exercise intensity. In practice this means the model may erroneously direct practitioners to over-exert patients during exercise rehabilitation sessions. We assess the impact of this further in Section \ref{sec:ProbPreds}.

There are differences in how the two models perform for each category. While the performance for observed rest values is similar, the smoothed model performs better for the low and medium observations, but sees a large drop in accuracy for high observations.  This is a natural result of the smoothing, which shifts high \(\dot VO_2\) values closer to the boundary between medium and high observations. In this region smaller errors in the predicted \(\dot VO_2\) values can result in incorrect classifications, impacting the accuracy of the model, while, for the same error, classifications in the raw model would be unaffected.

\begin{figure}[!ht]
\includegraphics[width = 16cm]{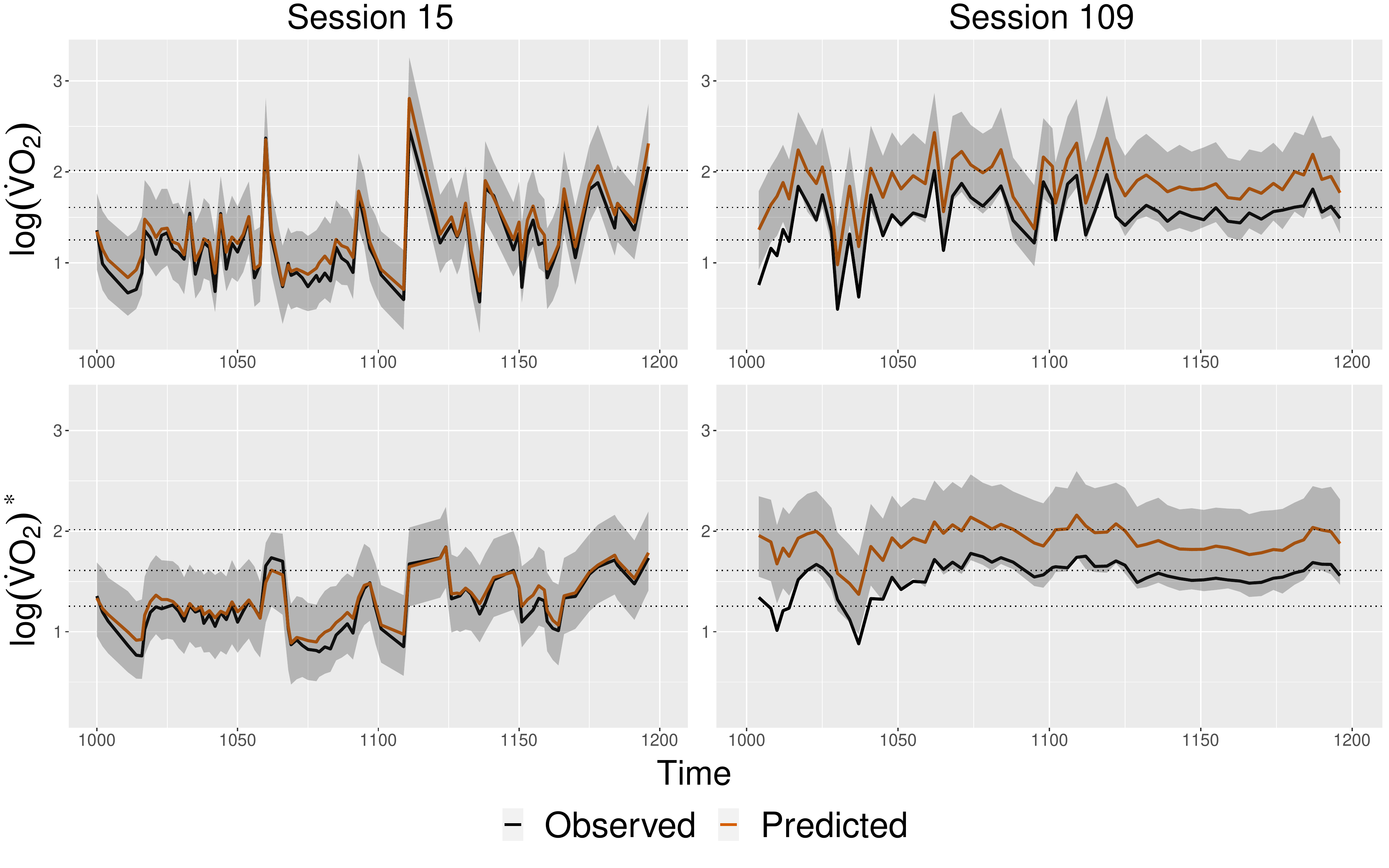}
\caption{Compares predicted (with 95\% credible intervals) and observed values of \(\dot VO_2\) for two example sessions, one where the model performs well (session 15) and one where it performs poorly (session 109). We use \(\log(\dot VO_2)^*\) to denote smoothed \(\dot VO_2\) values.}
\label{fig:4PredOT}
\end{figure}

To further understand where the model is and is not performing well we directly examined the predictions for \(\dot VO_2\). Figure \ref{fig:4PredOT} shows predicted \(\dot VO_2\) values over time for two sessions, for both models. In all cases the models predict the shape of the \(\dot VO_2\) curves extremely well (as expected from the posterior predictive checks). The plots for session 109 highlight, however, that for some sessions it is unable to quantify the scale of the curve, resulting in inaccurate classifications.  The magnitude of this error varies both between patients and between sessions for a single patient, which is to be expected given the high levels of between session heterogeneity noted in Section \ref{sec:Hierarchical}.

An alternative explanation for these poor predictions may be measurement error, either from the clinical side (e.g clinicians having to guess a patients weight), or from the technical side (e.g if the measurement device produces erroneous measurements).  This is supported by variations in model accuracy caused by session quality. Poor quality sessions on average had an accuracy of 48.5\% compared with 61.2\% for reasonable sessions and 65.1\% for good sessions, indicating that measurement error may be playing a role in impeding model performance. Note, however, that this decrease in average accuracy is primarily the result of an increase in variability of out of sample error for poorer quality sessions, rather than a universal drop in accuracy.

A further practical implication highlighted by Figure \ref{fig:4PredOT} is the stability provided by the smoothed model. When spikes in \(\dot VO_2\) values occur the raw model is able to more accurately predict \(\dot VO_2\), but these sudden changes in the probabilities and overall classifications that the model provides may not be as useful in practice, compared to the more stable classifications provided by the smoothed model.

\subsection{Probabilistic predictions}
\label{sec:ProbPreds}
Until now, we have primarily assessed the single classifications provided by the model. Probabilistic statements about the classifications are also given, however, and these may carry important information for clinicians.

\begin{figure}
\includegraphics[width = 16cm]{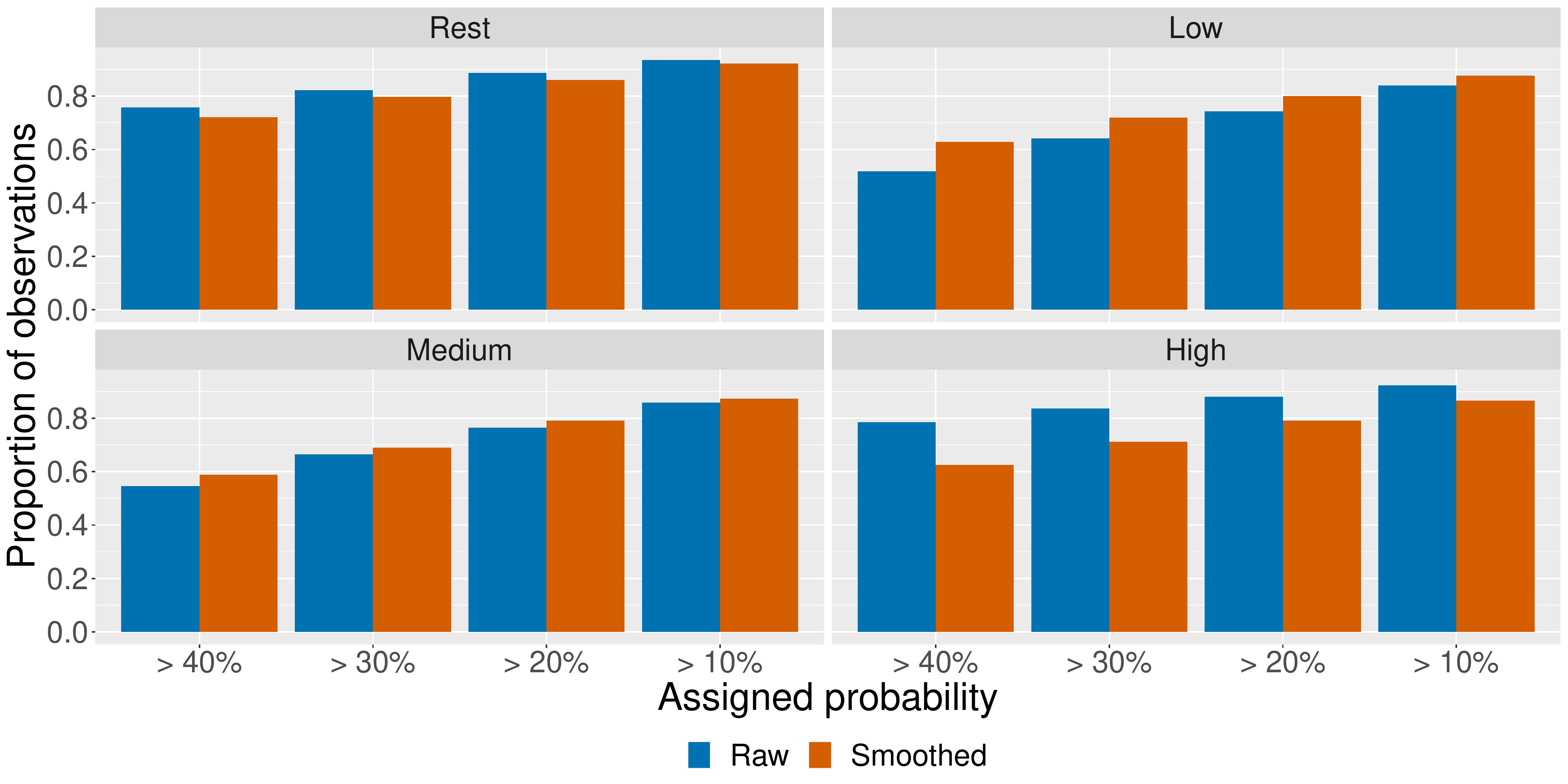}
\caption{Graphs by observed exercise intensity category for the raw (blue) and smoothed (red) model, showing the proportion of observations for which the model considers the correct classification plausible, based on different plausibility thresholds. }
\label{fig:4ProbAccount}
\end{figure}

In Figure \ref{fig:4ProbAccount} we assess the usefulness of this information by showing the number of observations within each category that the model assigns probabilities above a certain threshold. Heuristically, this is a way of assessing when the model considers the correct classification to be ``plausible", based on a set probability threshold for plausibility.

This is a weaker condition than seen in Table \ref{tab:ConfMatrices} and we naturally see a larger number of observations assigned as plausible. For example, at the 20\% threshold for plausible observations, 82.2\% and 82.8\% of observations are plausible in the raw and smoothed data models respectively.
This benefit is most notable for the Low and Medium categories, where around 80\% of observations are assigned as plausible for the 20\% threshold.

As noted in Section \ref{sec:Val_Results} the model has a tendency to under-predict exercise intensity. This is most concerning when observed \(\dot VO_2\) is high, as this indicates that the exercise intensity level should be lowered to avoid over-exertion. To assess the impact of this feature, we can use the results of Figure \ref{fig:4ProbAccount}, to see how the model would perform in these scenarios.

To avoid risks of over-exertion, a clinician may in practice apply an informal decision rule to lower a patient's exercise load once the probability they are at high intensity crosses a certain threshold.

This aligns exactly with the information showcased in Figure \ref{fig:4ProbAccount}. At the 20\% threshold these results indicate that the model would be able to identify 88.0\% and 79.1\% of high values for the raw and smoothed models respectively, correctly guiding the clinician in the majority of cases. 

This is only one example of how the full information from this model may be utilised and in practice a variety of rules may be implemented. It does indicate, however, that the full potential of the model is only realised once probabilistic statements about classifications are fully utilised.

\section{Discussion}
\label{sec:Disc}

We have developed and internally validated a first of its kind prediction model for exercise intensity in mechanically ventilated intensive care patients, using Bayesian hierarchical modelling and covariates that are readily available in the majority of intensive care settings. 

The model revealed a high level of heterogeneity between exercise rehabilitation sessions and the resulting classifications based on maximising posterior probability therefore had an accuracy of 60.0\%.  In Section \ref{sec:ProbPreds}, however, we showed how the usefulness of the model dramatically increases in practical situations when the full information from the model is utilised, in the form of probabilistic statements about classifications. In our example, the model was able to guide clinicians to make the correct decision 88.0\% of the time.

We have presented two feasible models using raw and smoothed physiological covariates. While the smoothed model provides more stable predictions and has a higher accuracy, the raw data model is noticeably better at classifying high intensity observations, which are particularly clinically important in ensuring that patients are not over-exerted. Further work is needed to decide which of the two models would be best to implement in practice.

\subsection{Future directions}

There is more work to be done in the study of exercise rehabilitation underpinning the model. In particular the categories that we have used to classify exercise intensity are principled and based on expert knowledge, but have been created specifically for this work and do not appear elsewhere in the literature. Future work could focus on refining and optimising them to ensure that clinicians are receiving as much information as possible about their patients whilst also maximising the predictive ability of the model.

As with much statistical work, the collection of further data would see several benefits. Given the small number of patients in our sample the model has only been trained on patients over 50 and has limited applicability to younger patients. Furthermore, the techniques used by \cite{Black2020} are cutting-edge, but not without issue. Techniques for measuring \(\dot VO_2\) can lack precision and are rarely applied to mechanically ventilated patients.\citep{Black2015}  Increased application to mechanically ventilated patients would therefore not only increase the amount of data available, but with refinement of techniques and improvement in technology could see an improvement in data quality. This would have the joint impact of removing any bias present in the posterior estimates, allowing us to better assess the accuracy of the model and expand the model's applicability to younger patients.

Additionally, more data would allow us to expand the applicability of the model to different populations. Importantly, \(\dot VO_2\) data for COVID-19 patients and replication of the work done here could be crucial.  Given the large number of patients who continue to be mechanically ventilated as a result of COVID-19 and the widely documented impact long-COVID has had on large proportions of those patients,\citep{Venkatesan2021} a similar prediction model developed specifically for that population could be hugely impactful.

The ultimate aim of the modeling work presented here is the creation of a practical tool that can be widely implemented in critical care environments. If this model can then be externally validated, it could provide the basis for a crucial tool for critical care clinicians. Once implemented, the ability to provide clinicians with accurate predictions of a patient's exercise intensity, in the easily interpretable form of exercise intensity categories, at the very least will allow for exercise rehabilitation to be rigorously tested in randomised clinical trials. If shown to be effective this could then be the first step towards patients receiving personalised exercise rehabilitation regimes and significant improvements in post-ICU outcomes.


\section*{Acknowledgments}
LH acknowledges support from EPSRC grant number EP/W523835/1 during part of this work. The authors would like to thank the UCL CHIMERA group, and in particular Prof. Christina Pagel and Prof. Rebecca Shipley for their invaluable comments and feedback.

\subsection*{Financial disclosure}

None reported.

\subsection*{Conflict of interest}

The authors declare no potential conflict of interests.

\section*{Data availability statement}

The data that support the findings of this study are available on request from the corresponding author. The data are not publicly available due to privacy or ethical restrictions.
\bibliographystyle{unsrtnat}
\bibliography{QEIICU,R}

\appendix

\section{Existing relationships with physiological covariates}
\label{sec:PhysRels}

While \(\dot VO_2\) is not readily available in the majority of critical care environments, there are several other physiological quantities that are available and have known relationships with \(\dot VO_2\). In this section we introduce these variables, and motivate the ability to generate a predictive model using them. Physiological covariates available in our data are outlined in Table \ref{tab:Vars}.

\subsection{Oxygen uptake efficiency slope}
\label{sec:OUES}

The relationship that has received the most attention in previous research is that between \(\dot VO_2\) and minute ventilation (\(\dot V_E\)) \citep{Ramos2013}. An existing relationship between the two quantities is the Oxygen Uptake Efficiency Slope (OUES) \citep{Baba1996}, where oxygen uptake efficiency is the amount of gas an individual needs to breathe in and out in order to utilise a given volume of oxygen. The OUES is defined by the equation
\begin{equation}
\label{eqn:OUES}
\dot VO_2 = a \times \log(\dot V_E) + b,
\end{equation}
and is simply the change in \(\dot VO_2\) as \(\log(\dot V_E)\) increases.  The OUES is significant in the exercise intensity literature as a measure of physical fitness independent of a patient's motivation to exercise \citep{Baba1996}. This can also be extended to further physiological covariates as minute ventilation is defined as the product of tidal volume (\(\dot V_T\)) and respiratory rate (\(RR\)).

\subsection{\(\dot V_E\) and \(\dot VCO_2\)}
\label{sec:VEVCO2}

A second relationship of interest is that between \(\dot V_E\) and \(\dot VCO_2\) defined as,
\begin{equation}
\label{eqn:2VEVCO2}
\dot V_E = \frac{863 \times \dot VCO_2}{P_aCO_2 \times (1 - \frac{\dot V_D}{\dot V_T})} , \\
\end{equation}
where
\begin{equation}
\label{eqn:2VD}
\dot V_D = \frac{P_aCO_2 - P_{ET}CO_2}{P_aCO_2}  ,\\
\end{equation}
and \(\dot VO_2\) is related to \(\dot VCO_2\) through the respiratory quotient,
\begin{equation}
\label{eqn:2RQ}
\dot VO_2 = \frac{\dot VCO_2}{RQ} .
\end{equation}
Here \(RQ\) typically takes values between 0.7 and 1.2.

There is therefore a relationship between \(\dot V_E\) and \(\dot VO_2\) via \(\dot VCO_2\), which is determined by \(\dot V_T\), \(P_aCO_2\), \(\dot V_D\), \(P_{ET}CO_2\) and \(RQ\).

Both these relationships are promising, and provide key motivation for why we may be able to model \(\dot VO_2\) from covariates available in the critical care context. In both cases, however, they have been shown to only apply in specific instances, with the OUES only holding below the anaerobic threshold, and the relationship in \eqref{eqn:2VEVCO2} only holding between the initial kinetic phase of exercise and the lactate threshold \citep{Ward2007}.

\begin{table}[!ht]
\bgroup
\def\arraystretch{1.2}
\begin{tabular}{|p{2cm}|p{14cm}|}
\hline
\textbf{Variable}                        & \textbf{Description}         \\
\hline\hline
\(\dot VO_2\)                     &     Volume of oxygen uptake. Values are not readily available for most intensive care patients and are measured through BBGEA.    \\
\hline
\( V_T\)            &   Tidal volume, the amount of air moved in and out of the lungs during a single breath.           \\
\hline
\(RR\)       &      Respiratory rate, the number of breaths per minute.          \\
\hline
\(\dot V_E\)          &        Minute ventilation, the amount of air breathed per minute, defined as the product of tidal volume and respiratory rate.      \\
\hline
\(P_{ET}CO_2\)                &      Amount of CO\textsubscript{2} in exhaled air on a breath-by-breath basis.            \\
\hline
\(P_aCO_2\)          &          Partial pressure of Carbon dioxide.                   \\
\hline
\(\dot V_D\)                    &  Dead space volume, the volume of air inhaled that does not take part in gas exchange.                 \\
\hline
\(\dot VCO_2\)                               &          Volume of exhaled carbon dioxide.       \\
\hline
RQ                            &          Respiratory quotient, the ratio of \(\dot VO_2\)     to \(\dot VCO_2\)      (see equation \ref{eqn:2RQ}).    \\
\hline
\end{tabular}
\caption{Physiological variables used in the above relationships.}
\label{tab:Vars}
\egroup
\end{table}
\clearpage

\section{Further exploratory analysis}

\begin{figure}[!ht]
\includegraphics[width = 16cm]{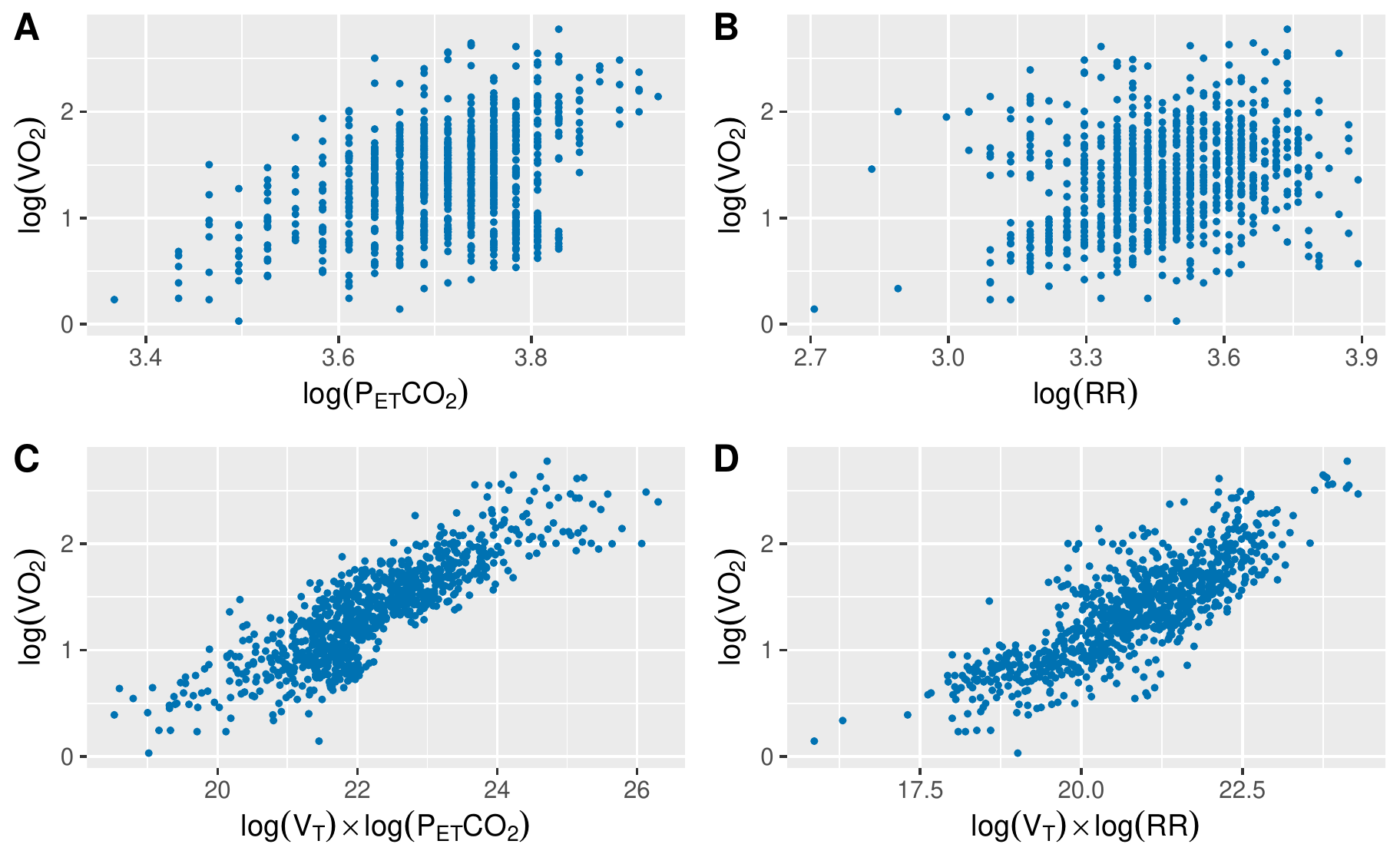}
\caption{For test 15: (A) The log-log relationship between \(\dot VO_2\) and \(P_{ET}CO_2\). (B) The log-log relationship between \(\dot VO_2\) and Respiratory Rate (\(RR\)). (C) The relationship between \(\dot VO_2\) and the product of \(\log(\dot V_T)\) and \(\log(P_{ET}CO_2)\). (D) The relationship between  \(\dot VO_2\) and the product of \(\log(\dot V_T)\) and \(\log(RR)\).}
\end{figure}

\begin{figure}[!ht]
\includegraphics[width = 16cm]{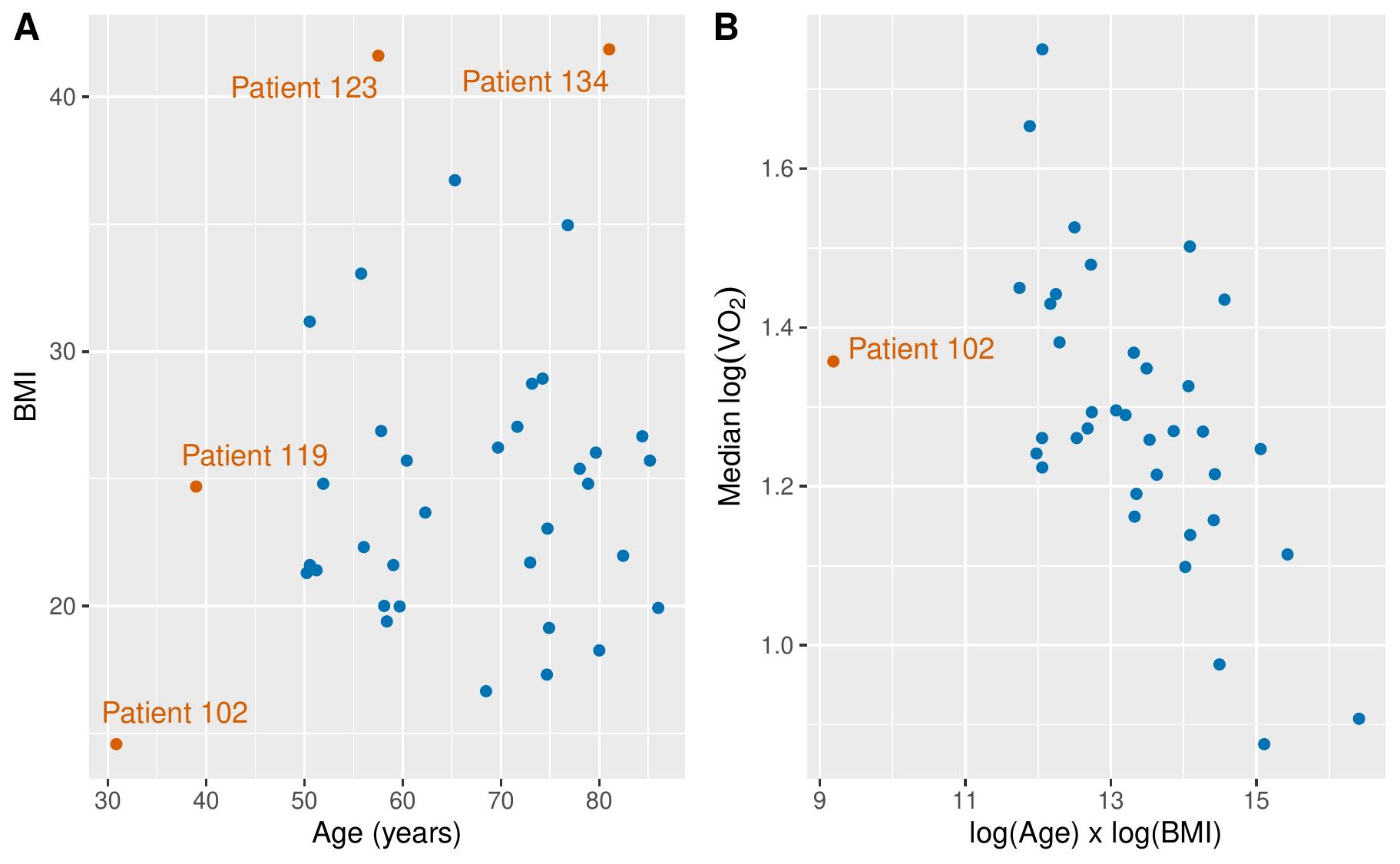}
\caption{Plots showing the relationship between patients' age, BMI and median \(\dot VO_2\) levels. (A) shows age in years against BMI, with patients 119 highlighted as an outlier due to their age, patients 123 and 134 highlighted as outliers due to their BMI, and patient 102 highlighted as an outlier due to their age and BMI. (B) shows a potential interaction between effect between \(\log(age)\) and \(\log(BMI)\), with patient 102 highlighted again as an outlier.}
\end{figure}

\clearpage

\section{Directed acylic graph for the model in Section 3}

\begin{figure}[h]
    \centering
        \includegraphics[clip, trim=4cm 11cm 4cm 4cm, width=1.00\textwidth]{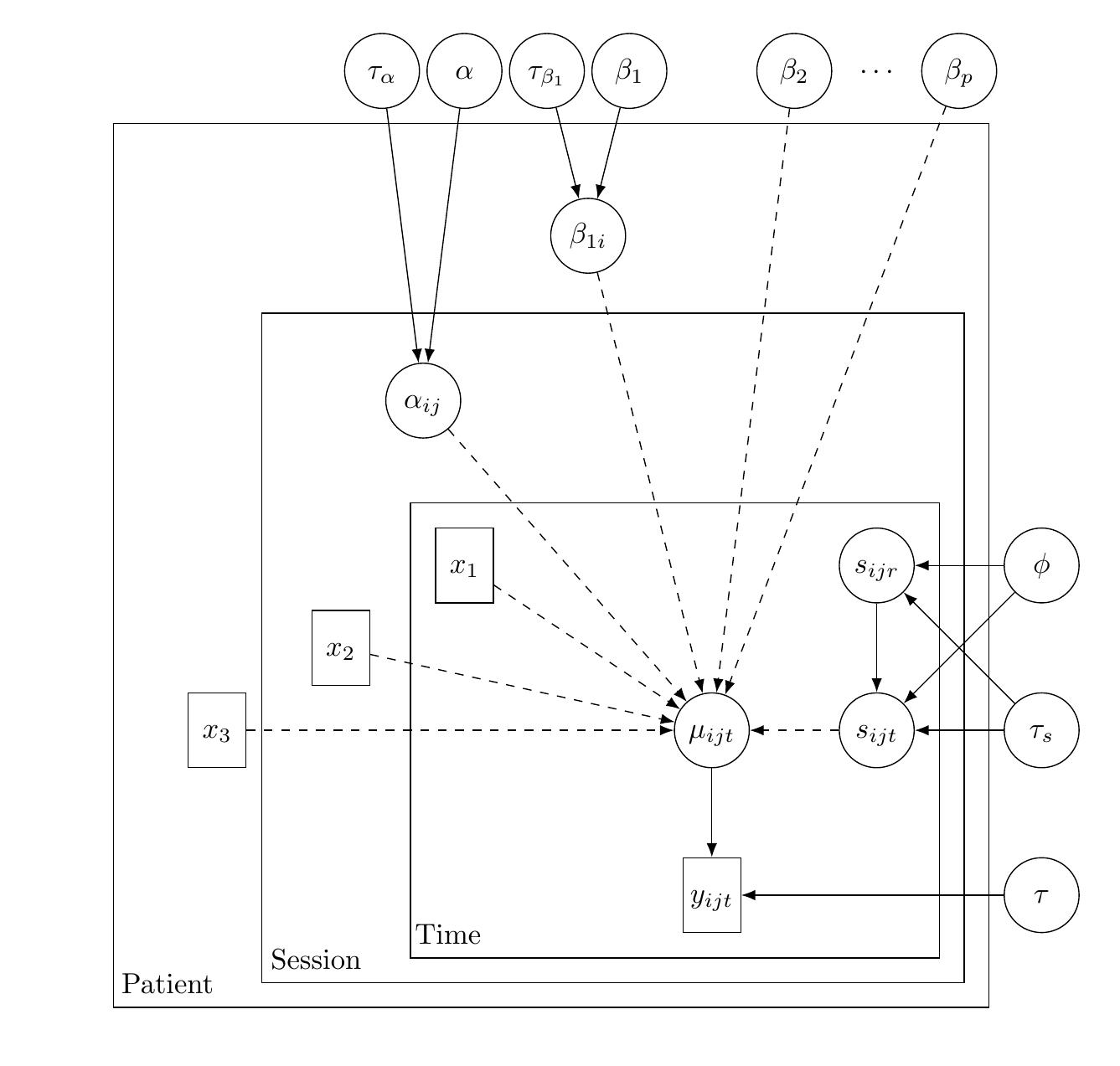}
    \caption{Directed Acyclic Graph for the model presented in Section \ref{sec:TheMod}. Solid lines indicate distributional dependencies and dashed lines indicate deterministic dependency. Covariates are indicated by \(x_k\)'s with \(x_1\) indicating \(V_T, P_{ET}CO_2\) and respiratory rate, \(x_2\) indicating SOFA and \(x_3\) indicating age, BMI and GPPAQ. The time of the previous observation is denoted by \(r\). If \(t\) is the first observation time then \(s_{ijt} \sim \text{Normal}(0,\tau_s)\). }
\end{figure}
\clearpage
\section{Visualisation of heterogeneity in the \(V_T\) coefficient}

\begin{figure}[!ht]
\includegraphics[width = 16cm]{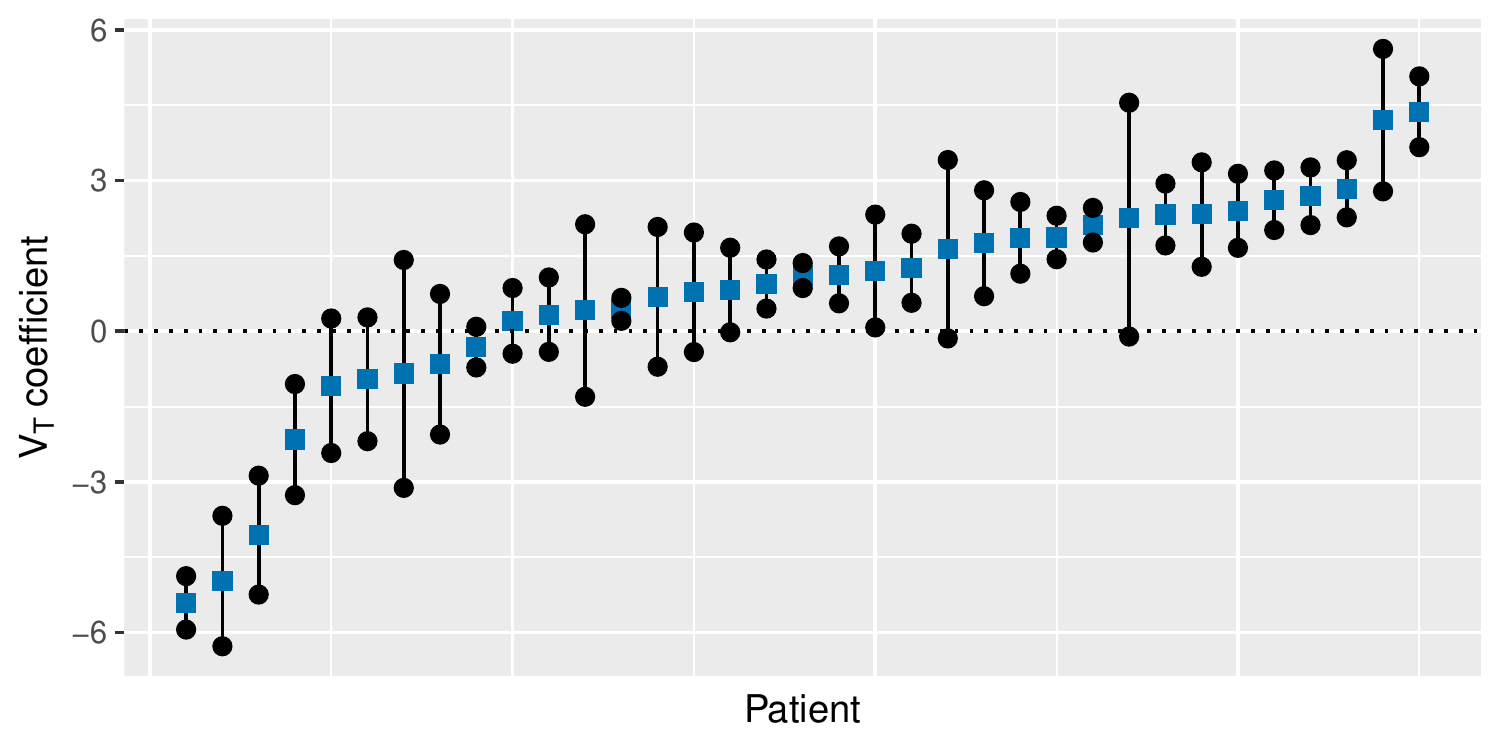}
\caption{Plots showing \(\dot VO_2\), \(\dot V_E\) and \(V_T\) over time for session 73, with disruption caused by a cough circled.}
\end{figure}
\clearpage
\section{Posterior summaries for alternative models}

\begin{center}
\begin{table}[!ht]%
\centering
\caption{Posterior summaries for the fixed effects coefficients and hyperparameters, with mean, standard deviation (SD) and a 95\% central posterior interval for the model using smoothed data.($\dagger$  Fixed effects with 95\% posterior credible intervals that do not contain 0).\label{tab:4Posts}}%
\begin{tabular*}{350pt}{@{\extracolsep\fill}lrrrr@{\extracolsep\fill}}
\toprule
 & \textbf{Mean}  & \textbf{SD}  & \textbf{2.5\%}  & \textbf{97.5\%} \\
\midrule
Intercept\(\dagger\) & 1.32 & 0.08 & 1.17  & 1.46 \\ 
  \(\log ( V_T)\dagger\) & 1.49 & 0.03 & 1.44  & 1.54 \\ 
  \(\log (P_{ET}CO_2)\dagger\) & 1.65 & 0.01 & 1.63  & 1.66  \\ 
  \(\log (RR)\dagger\) & 1.05 & \(<\)0.01 & 1.05  & 1.06   \\ 
  \(\log ( V_T) \times \log (P_{ET}CO_2)\dagger\) & -0.19 & 0.02 & -0.22  & -0.15 \\ 
  \(\log ( V_T) \times \log (RR)\dagger\) & -0.06 & \(<\)0.01 & -0.06  & -0.05   \\ 
  SOFA & -0.01 & 0.02 & -0.04  & 0.03   \\ 
  GPPAQ = 2 & 0.01 & 0.04 & -0.08  & 0.09 \\ 
  GPPAQ = 3 & 0.02 & 0.04 & -0.07  & 0.10   \\ 
  GPPAQ = 4\(\dagger\) & 0.25 & 0.06 & 0.13  & 0.37  \\ 
  Sex & -0.09 & 0.04 & -0.17 & -0.02  \\ 
  \(\log(age)\dagger\) & 0.28 & 0.10 & 0.08  & 0.47  \\ 
  \(\log(BMI)\dagger\) & -1.02 & 0.07 & -1.17  & -0.88   \\ 
  \(\log(age) \times \log(BMI)\dagger\) & -1.89 & 0.53 & -2.93  & -0.86  \\ 
  \midrule
  \(\tau_\alpha\) & 52.82 & 3.07 & 46.64  & 58.64  \\ 
  \(\tau_\beta\) & 51.35 & 8.71 & 34.82  & 68.70  \\ 
  O-U \(\tau\) & 53.75 & 0.79 & 52.14  & 55.24  \\ 
  O-U \(\phi\) & 0.03 & \(<\)0.01 & 0.03  & 0.03 \\ 
\bottomrule
\end{tabular*}

\end{table}
\end{center}

\begin{center}
\begin{table}[!ht]%
\centering
\caption{Posterior summaries for the fixed effects coefficients and hyperparameters, with mean, standard deviation (SD) and a 95\% central posterior interval for the model fitted with only good and reasonable quality tests.($\dagger$  Fixed effects with 95\% posterior credible intervals that do not contain 0).\label{tab:4Posts}}%
\begin{tabular*}{350pt}{@{\extracolsep\fill}lrrrr@{\extracolsep\fill}}
\toprule
 & \textbf{Mean}  & \textbf{SD}  & \textbf{2.5\%}  & \textbf{97.5\%} \\
\midrule
Intercept\(\dagger\) & 1.31 & 0.12 & 1.07  & 1.54 \\ 
  \(\log ( V_T)\dagger\) & 1.56 & 0.03 & 1.50  & 1.63 \\ 
  \(\log (P_{ET}CO_2)\dagger\) & 1.91 & 0.01 & 1.91  & 1.93  \\ 
  \(\log (RR)\dagger\) & 1.11 & \(<\)0.01 & 1.10  & 1.11   \\ 
  \(\log ( V_T) \times \log (P_{ET}CO_2)\dagger\) & -0.25 & 0.01 & -0.27  & -0.22 \\ 
  \(\log ( V_T) \times \log (RR)\dagger\) & 0.33 & \(<\)0.01 & 0.32  & 0.34   \\ 
  SOFA & -0.01 & 0.02 & -0.06  & 0.04   \\ 
  GPPAQ = 2 & 0.01 & 0.07 & -0.13  & 0.14 \\ 
  GPPAQ = 3 & 0.01 & 0.07 & -0.13  & 0.13   \\ 
  GPPAQ = 4\(\dagger\) & 0.34 & 0.10 & 0.14  & 0.55  \\ 
  Sex & -0.03 & 0.06 & -0.14 & 0.09  \\ 
  \(\log(age)\dagger\) & 0.45 & 0.16 & 0.12  & 0.76  \\ 
  \(\log(BMI)\dagger\) & -1.07 & 0.11 & -1.29  & -0.85   \\ 
  \(\log(age) \times \log(BMI)\dagger\) & -1.91 & 0.81 & -3.51  & -0.31  \\ 
  \midrule
  \(\tau_\alpha\) & 20.99 & 2.69 & 17.00  & 27.44  \\ 
  \(\tau_\beta\) & 31.69 & 8.11 & 16.36  & 47.64  \\ 
  O-U \(\tau\) & 48.59 & 0.56 & 47.47  & 49.68  \\ 
  O-U \(\phi\) & 0.09 & \(<\)0.01 & 0.09  & 0.10 \\ 
\bottomrule
\end{tabular*}
\end{table}
\end{center}

\begin{center}
\begin{table}[!ht]%
\centering
\caption{Posterior summaries for the fixed effects coefficients and hyperparameters, with mean, standard deviation (SD) and a 95\% central posterior interval, for the model fitted with only good quality tests.($\dagger$  Fixed effects with 95\% posterior credible intervals that do not contain 0).\label{tab:4Posts}}%
\begin{tabular*}{350pt}{@{\extracolsep\fill}lrrrr@{\extracolsep\fill}}
\toprule
 & \textbf{Mean}  & \textbf{SD}  & \textbf{2.5\%}  & \textbf{97.5\%} \\
\midrule
Intercept\(\dagger\) & 1.26 & 0.14 & 0.99  & 1.54 \\ 
  \(\log ( V_T)\dagger\) & 1.59 & 0.03 & 1.53  & 1.66 \\ 
  \(\log (P_{ET}CO_2)\dagger\) & 1.85 & 0.01 & 1.83  & 1.87  \\ 
  \(\log (RR)\dagger\) & 1.13 & \(<\)0.01 & 1.12  & 1.13   \\ 
  \(\log ( V_T) \times \log (P_{ET}CO_2)\dagger\) & -0.39 & 0.02 & -0.43  & -0.35 \\ 
  \(\log ( V_T) \times \log (RR)\dagger\) & 0.26 & \(<\)0.01 & 0.25  & 0.28   \\ 
  SOFA & -0.02 & 0.03 & -0.08  & 0.03   \\ 
  GPPAQ = 2 & -0.04 & 0.08 & -0.20  & 0.11 \\ 
  GPPAQ = 3 & 0.01 & 0.08 & -0.14  & 0.16   \\ 
  GPPAQ = 4 & 0.12 & 0.12 & -0.12  & 0.36  \\ 
  Sex & 0.10 & 0.07 & -0.04 & 0.24  \\ 
  \(\log(age)\dagger\) & 0.54 & 0.24 & 0.06  & 1.02  \\ 
  \(\log(BMI)\dagger\) & -0.85 & 0.15 & -1.14  & -0.56   \\ 
  \(\log(age) \times \log(BMI)\) & -1.52 & 1.10 & -3.69  & 0.65  \\ 
  \midrule
  \(\tau_\alpha\) & 51.75 & 7.65 & 36.98  & 66.94  \\ 
  \(\tau_\beta\) & 50.30 & 7.63 & 35.23  & 64.89  \\ 
  O-U \(\tau\) & 51.5 & 1.19 & 48.92  & 53.53  \\ 
  O-U \(\phi\) & 0.07 & \(<\)0.01 & 0.07  & 0.08 \\ 
\bottomrule
\end{tabular*}
\end{table}
\end{center}

\clearpage

\end{document}